\documentclass{article}
\usepackage[a4paper, total={6in, 8in}]{geometry}
\usepackage{cite}
\usepackage{amsmath,amssymb,amsfonts}
\usepackage{algorithmic}
\usepackage{graphicx}
\usepackage{textcomp}
\usepackage{multirow}
\usepackage{caption}
\usepackage{subcaption}
\usepackage{hyperref}

\title{Evaluation of large language models for assessing code maintainability}

\author{Marc Dillmann, Julien Siebert, Adam Trendowicz}

\begin{document}

\maketitle

\begin{abstract}
    \textbf{Background}: Increased availability of open-source software repositories and recent advances in code analysis using large language models (LLMs) has triggered a wave of new work to automate software engineering tasks that were previously very difficult to automate (e.g. , generating code from text). \\
    \textbf{Objective}: In this paper, we investigate a recent line of work that hypothesises that comparing the probability of code generated by LLMs (counterfactual) with the probability the current code would have had can indicate potential quality problems. \\
    \textbf{Method}: We investigate the association between the cross-entropy of code generated by ten different models (based on GPT2 and Llama2) and the following quality aspects: readability, understandability, complexity, modularisation, and overall maintainability assessed by experts and available in an benchmark dataset. \\
    \textbf{Result}: Our results show that, controlling for the number of logical lines of codes (LLOC), cross-entropy computed by LLMs is indeed a predictor of maintainability on a class level (the higher the cross-entropy the lower the maintainability). However, this relation is reversed when one does not control for LLOC (e.g., comparing small classes with longer ones). Furthermore, while the complexity of LLMs affects the range of cross-entropy (smaller models tend to have a wider range of cross-entropy), this plays a significant role in predicting maintainability aspects.\\
    \textbf{Limitation:} Our study limits itself on ten different pretrained models (based on GPT2 and Llama2) and on maintainability aspects collected by Schnappinger \textit{et al.} \cite{schnappinger_dataset_2020}.\\
    \textbf{Conclusion}: When controlling for logical lines of code (LLOC), cross-entropy is a predictor of maintainability. However, while related work has shown the potential usefulness of cross-entropy at the level of tokens or short sequences, at the class level this criterion alone may prove insufficient to predict maintainability (LLOC is still a stronger predictor and should be controlled for), and further research is needed to make best use of this information in practice.
\end{abstract}

\section{Introduction}\label{sec:introduction}
The availability of software repositories (e.g., Github) combined with new advances in code analysis based on large language models (LLMs) has triggered a wave of new work to automate tasks that were previously very difficult to automate, such as generating code from natural language, generating documentation from code, extracting requirements from text, etc. \cite{allamanis2019survey,yang2022survey,wu2022survey}. Some of these models are already available to developers and used in tools such as Microsoft's Github Copilot, Salesforce's CodeGen, Meta's Code Llama, Tabnine or Amazon Web Service Codewhisperer, etc. These new capabilities have also led to research focusing on how such models can be used for software quality assessment. Methods based on deep neural networks (specifically LLMs) offer many potential benefits compared to approaches based on structural properties of software artifacts, such as code \cite{briand2000exploring}. Example benefit include their direct application on unstructured data, such as text (code) or image (architecture) without prior extraction of structured features such as code metrics. This is important because not only does the collection of structured metrics require additional work and is not always feasible \cite{trendowicz2021data}, but also the usefulness of software metrics as indicators of specific software characteristics has raised much controversy in the software engineering community (e.g. \cite{shepperd1988critique,fenton1999critique}).

Application of LLMs for the purpose of software quality assessment is however not free from challenges. While most research focuses on quality aspects that are more amenable to quantification, such as bug prediction or vulnerability prediction, we see the emergence of work targeting quality aspects that are more difficult to quantify, such as maintainability\footnote{See the details of our previous systematic literature recherche in \url{https://github.com/marcdillmann/DeepCodeMaintainability}}. Further research is thus needed to evaluate the applicability of these models in the context of software quality assessment.

In this paper we empirically evaluate how LLMs can be used to assess code maintainability on a class level. We investigate the assumption that the probability of generated code can be an indicator of code maintainability \cite{sengamedu_neural_2022}. In other words, LLMs are assumed to be oracles who have learned good programming practices, and their generated code should reflect such practices. To do this, we investigate the association between the cross-entropy of code generated by 10 different models (based on GPT2 and Llama2) and the following quality aspects: readability, understandability, complexity, modularisation, and overall maintainability being measured by experts and available from a benchmark dataset\footnote{This dataset is available on GitHub: \url{https://github.com/simonzachau/SWQD-predict-software-maintainability}} \cite{schnappinger_defining_2020,schnappinger_dataset_2020}.

Our results show that, controlling for the number of logical lines of codes (LLOC), cross-entropy computed by LLMs is indeed a predictor of maintainability on a class level (the higher the cross-entropy the lower the maintainability). However, this relation is reversed when one does not control for LLOC (e.g., comparing small classes with longer ones). Furthermore, the complexity of LLMs (the number of trainable parameters) does not seem to play a significant role for this task. 

This paper is organised as follow. An overview of the related work is provided in section \ref{sec:related-work}. The details of our experimental setup and the results are given in sections \ref{sec:methods} and \ref{sec:results}. Section \ref{sec:disccusions} summarizes and discussed this work.

\section{Related Work}\label{sec:related-work}
Schnappinger \textit{et al.} \cite{schnappinger_preliminary_2022} investigate the performance of text-based and image-based machine learning (ML) algorithms for predicting code readability, understandability, and complexity. In a previous work, the authors have collected and curated a data set of 304 Java classes together with expert opinions on their maintainability aspects (readability, understandability, complexity, modularisation, and overall maintainability) \cite{schnappinger_defining_2020,schnappinger_dataset_2020}. This data set is used as ground truth for training and testing several ML models. The results show that classical ML classifiers (such as SVM or Naive Bayes) achieve similar performance as human experts, but that deep neural network models (such as AxelNet and Bert) underperform drastically. 

Hu \textit{et al.} propose a deep learning model to predict the maintainability of source code \cite{hu_maintainability_2023}. Their approach relies on lexical features (i.e., text), structural representation of the code (i.e., abstract syntax tree), and code metrics. While their training data is collected in an automated way, their evaluation data set is manually collected and curated.
Their results show good performance on the task at hand.

Both works rely on training models for the specific task of predicting code maintainability aspects and require, at least, a manually curated evaluation data set. The next paper focuses on using pretrained LLMs to assess maintainability.

Sengamedu and Zhao \cite{sengamedu_neural_2022} introduce a novel approach for detecting code quality issues using LLMs. The underlying hypothesis is that these models inherently contain a great deal of information about the way code should be written, and that comparing the probability of the generated code with the current code can indicate potential quality problems. The authors define four types of quality problems and how LLMs can help detect them.
\textbf{Token-level errors} can be detected when the current token (written by a human) has a lower probability compared to other tokens being generated by the LLM.
\textbf{Unnatural code} (i.e., code that is potentially less readable and understandable) can be detected when the language model finds a given code sequence surprising and assigns it a low probability or a high cross-entropy and perplexity.
\textbf{Repetitive code} can be detected when the cross-entropy of a sequence is very low.
\textbf{Long and complex code} can be detected using the log probability (LP) of a code sequence. A high LP indicates that the code snippet is long and unpredictable.

The use of pre-trained LLMs to assess code quality is still in its infancy. Despite promising results, the assumption that LLMs can capture good coding practices and thus be used to assess code quality has yet to be systematically investigated.

\section{Method}\label{sec:methods}
This paper focuses on the following research question:

\textbf{RQ}: \textit{At a class level, is the cross-entropy between the probability of the generated token and the actual code a good indicator of code maintainability, specifically its readability, understandability, complexity, and modularization?}

To answer this question, we perform an empirical study, in which we investigate correlation between the software source cross-entropy (between generated and actual code) and the code's maintainability indicators assessed by human experts. For this purpose, use the data set from Schnappinger \textit{et al.} \cite{schnappinger_defining_2020,schnappinger_dataset_2020} containing Java classes and the corresponding expert ratings of five dimensions (\textit{readability}, \textit{understandability}, \textit{complexity}, \textit{modularity}, and \textit{overall maintainability}). We employ ten different pretrained LLMs (see Table \ref{tab:models}) to generate code, based on the java class files available in the evaluation data. Next, we compute the cross-entropy for each java class file, based on the next-token probabilities. We also compute, for each class, the number of logical lines of code (LLOC). Finally, we investigate the relation between cross-entropy of the code generated and the maintainability values for each of the five dimensions. 

\begin{figure*}[!htbp]
    \centering
    \includegraphics[width=\textwidth]{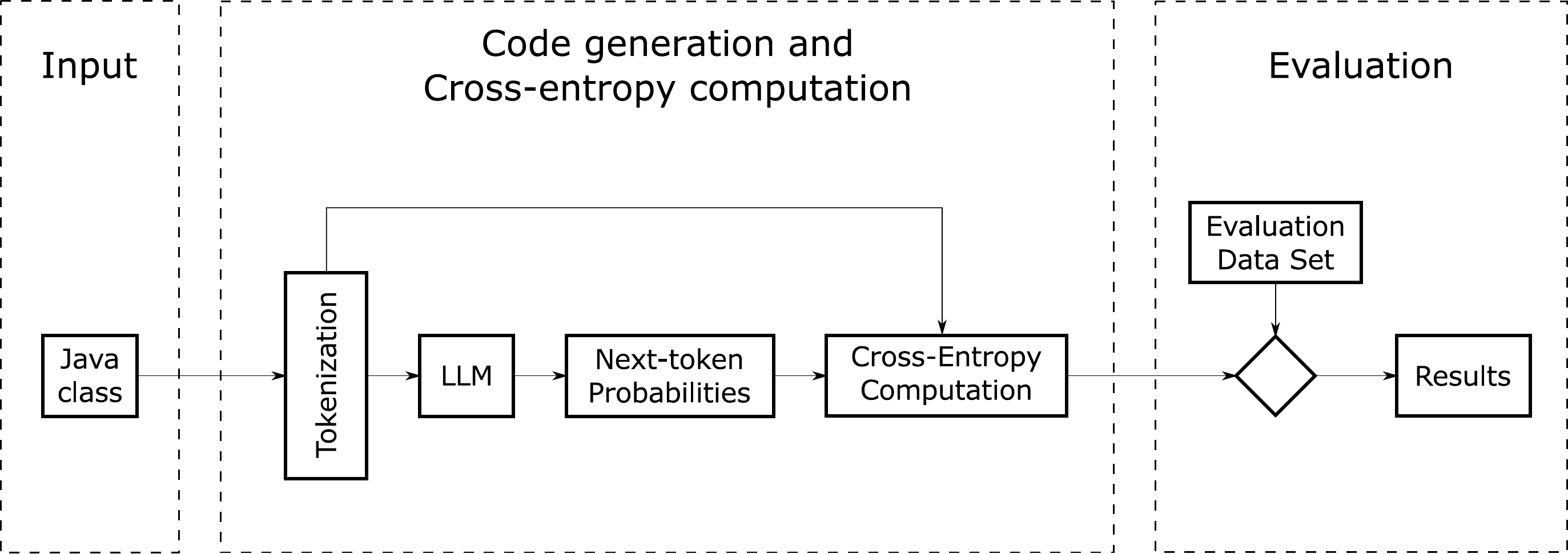}
    \caption{Overview of the experimental design}
    \label{fig:experimental-design}
\end{figure*}

Figure \ref{fig:experimental-design} illustrates our experimental design. The experimental code base is publicly available \footnote{\url{https://github.com/marcdillmann/DeepCodeMaintainability}} and relies upon the PyTorch\footnote{\url{https://pytorch.org}} and the Hugging Face\footnote{\url{https://huggingface.co}} transformers libraries. The next paragraphs provide more details about the implementation.

\begin{table*}[!htbp]
    \centering
    \small
    \begin{tabular}{ ccc }
        \hline
        \textbf{ID} & \textbf{Model Name} & \textbf{Max Input} \\
        \hline
        M1 & \href{https://huggingface.co/bigscience/bloomz-560m}{bigscience/bloomz-560m} & 1024  \\
        M2 & \href{https://huggingface.co/bigscience/bloomz-1b1}{bigscience/bloomz-1b1}  & 1536  \\
        M3 & \href{https://huggingface.co/bigscience/bloomz-1b7}{bigscience/bloomz-1b7}  & 2048  \\
        M4 & \href{https://huggingface.co/bigscience/bloomz-3b}{bigscience/bloomz-3b} & 2560  \\
        M5 & \href{https://huggingface.co/bigscience/bloomz-7b1}{bigscience/bloomz-7b1} & 4096  \\
        M6 & \href{https://huggingface.co/microsoft/CodeGPT-small-java-adaptedGPT2}{microsoft/CodeGPT-small-java-adaptedGPT2} & 1024  \\
        M7 & \href{https://huggingface.co/thmk/codegpt-java-10.2}{thmk/codegpt-java-10.2} & 1024  \\
        M8 & \href{https://huggingface.co/ammarnasr/codegen-350M-mono-java}{ammarnasr/codegen-350M-mono-java}  & 1024 \\
        M9 & \href{https://huggingface.co/codellama/CodeLlama-7b-hf}{codellama/CodeLlama-7b-hf}  & 16384  \\
        M10 & \href{https://huggingface.co/codellama/CodeLlama-13b-hf}{codellama/CodeLlama-13b-hf} & 16384  \\
        \hline
    \end{tabular}
    \caption{Overview of the selected models. The column "Model Name" refers to the identification string in Hugging Face. "Max Input" describes the maximum number of tokens that the model can handle as input.}
    \label{tab:models}
\end{table*}

\textbf{Preprocessing:} Before processing the Java code, we removed the comments and the empty lines. Next, each class is segmented  into chunks corresponding to the model's maximum permissible input size.

\textbf{Cross-entropy Computation:} 
Our approach requires some minor adjustments to accommodate our use of chunks. Specifically, we use PyTorch's CrossEntropyLoss() function to compute the cross-entropy for each chunk. We then adjust for the number of tokens within each segment. The cross-entropy for the whole class is an average of the cross-entropy values of the chunks, weighted by the number of tokens in each chunk. This method ensures that the impact, especially of the last --- potentially shorter --- chunk, is scaled.

\textbf{Evaluation data set:}
The evaluation data set provides, for each java class (304 in total), assessments of the considered five maintainability aspects which are given on a Likert scale for the following statements: Readability (Rd) - this code is easy to read; Understandability (Ud.) - the semantic meaning of this code is clear; Complexity (Cx.) - this code is complex; Modularity (md.) - this code should be broken into smaller pieces; Overall maintainability (Ov.) - overall, this code is maintainable. The 4-point Likert-scale consists of four values: P(strongly agree), P(weakly agree), P(weakly disagree), and P(strongly disagree), each corresponding to the probability of the expert's answer. For example, for the readability dimension (Rd.), P(strongly agree) denotes the probability that an expert will strongly agree with the statement that the code is easy to read. Table \ref{tab:eval_data} provides an overview of the first two lines of this data set. Note that this data set is imbalanced. For instance, for the dimension overall maintainability (Ov.), the probability of answering "strongly agree" is maximum for 174 out of 304 files, whereas it is maximum for "strongly disagree" for only 25 out of 304 files. Table \ref{tab:ntimes-proba-max} provides the number for all answers type and all five dimensions. Note that for both dimensions complexity (Cx.) and modularity (Md.) the question is negative and answering "strongly disagree" means that the code is maintainable (see \cite{schnappinger_defining_2020,schnappinger_dataset_2020} for details).

\begin{table*}[!htbp]
    \centering
    \footnotesize
    \begin{tabular}{|c|cccc|cccc|cccc|}
        \hline
         File & \multicolumn{4}{c|}{\textbf{Ov.}} & \multicolumn{4}{c|}{\textbf{Rd.}} & \multicolumn{4}{c|}{\textbf{Ud.}} \\ 
         \hline
         ActorEditorWindow.java & 0 & 0.24 & 0.76 & 0 & 0 & 0.85 & 0.15 & 0 & 0 & 0.17 & 0.83 & 0  \\
         \hline
         AnimationPreviewer.java & 0 & 0 & 0.34 & 0.66 & 0 & 0.04 & 0.95 & 0.01 & 0 & 0.02 & 0.97 & 0.01   \\ 
         \hline
    \end{tabular}\\
    \vspace{3mm}
    \begin{tabular}{|c|cccc|cccc|}
        \hline
         File & \multicolumn{4}{c|}{\textbf{Cx.}} & \multicolumn{4}{c|}{\textbf{Md.}} \\ 
         \hline
         ActorEditorWindow.java & 0.04 & 0.24 & 0.72 & 0 & 0 & 0.75 & 0.025 & 0 \\
         \hline
         AnimationPreviewer.java & 0 & 0.86 & 0.14 & 0 & 0.02 & 0.98 & 0 & 0  \\ 
         \hline
    \end{tabular}
    \caption{Overview of the first lines of the evaluation data set. The values are rounded to the second decimal place.}
    \label{tab:eval_data}
\end{table*}

\begin{table*}[!htbp]
\small
    \centering
    \begin{tabular}{ cccccc }
        \hline
         & \textbf{Ov.} & \textbf{Rd.} & \textbf{Ud.} & \textbf{Cx.} & \textbf{Md.}\\
        \hline
        strongly agree      & 174 & 183 & 157 &  22 &  29 \\
        weakly agree        &  64 &  79 &  76 &  41 &  31 \\
        weakly disagree     &  41 &  38 &  51 &  60 &  49 \\
        strongly disagree   &  25 &   4 &  20 & 181 & 195 \\
        \hline
    \end{tabular}
    \caption{Number of time the probability a given answer (strongly agree, weakly agree, weakly disagree, strongly disagree) is maximum. Note that for almost all cases this corresponds to a probability being greater than or equal to 0.5.}
    \label{tab:ntimes-proba-max}
\end{table*}

\textbf{Computing LLOC} 
For each class, we compute LLOC by simply counting the remaining lines of code after comments and empty lines have been removed. 

\textbf{Analysis methods}
We evaluated the task of predicting whether an expert would most likely strongly agree (i.e., P(strongly agree) > 0.5) for the Overall maintainability (Ov.), Readability (Rd.), and Understandability (Ud.) aspects and strongly disagree (i.e., P(strongly disagree) > 0.5) for the Complexity (Cx.) and Modularity (Md.) aspects. We examine different classifiers (C1: logistic regression, C2: random forest, and C3: support vector machine) with either LLOC as a single feature (baseline),  cross-entropy as a single feature, or cross-entropy and LLOC as features. Features are all first transformed using a log scale and standardized (mean = 0, std = 1). Note that this is a binary classification problem that is not perfectly balanced (see Table \ref{tab:number-instances-labels}). For Ov., Rd., and Ud., P(strongly agree) $\leq$ 0.5 is class 0 and P(strongly agree) > 0.5 is class 1. For Cx. and Md., P(strongly disagree) $\leq$ 0.5 is class 0 and P(strongly disagree) > 0.5 is class 1. We computed the mean value for each of the following metrics: accuracy (Acc), f1-score (F1), area under the receiver operator curve (ROC), and Matthew Correlation Coefficient (MCC), all measured on the test data set from a 10-fold cross validation. We did not perform any hyperparameters optimization. We use the sklearn implementation with default parameters except for logistic regression where we use \textit{penalty=False} and \textit{fit\_intercept=True} (which is equivalent to statsmodels logistic regression implementation with an intercept).

\textbf{Infrastructure \& Execution:} Our initial experiments leveraged an Nvidia RTX 3080 GPU with 10GB VRAM. Encoutered memory constraints, especially with larger models, required later a transition to a GPU cluster housed on a cloud computing platform. Most models, barring the Code Llama variants, were evaluated on a server equipped with a Nvidia RTX A6000 (40GB VRAM). The resource-intensive Code Llama models demanded even more capacity, prompting us to use a Nvidia A100 (80GB VRAM). See Table \ref{tab:runtime} for details.

\begin{table*}[!htbp]
    \centering
    \small
    \begin{tabular}{ cccc }
        \hline
        \textbf{ID} & \textbf{Model} & \textbf{Hardware (Nvidia)} & \textbf{Runtime} \\
        \hline
        M1 & Bloomz-560m &  RTX A6000 (40GB VRAM) & 00:50 \\
        M2 & Bloomz-1b1 &  RTX A6000 (40GB VRAM) &01:21  \\
        M3 & Bloomz-1b7 &  RTX A6000 (40GB VRAM) &02:05\\
        M4 & Bloomz-3b &  RTX A6000 (40GB VRAM) & 03:52 \\
        M5 & Bloomz-7b1 &  RTX A6000 (40GB VRAM) & 08:00 \\
        M6 & CodeGPT-small-java &  RTX A6000 (40GB VRAM) & 00:16  \\
        M7 & CodeGPT-java-10.2 &  RTX A6000 (40GB VRAM) & 00:14 \\
        M8 & codegen-350m &  RTX A6000 (40GB VRAM) & 00:49 \\
        M9 & CodeLlama-7b &  A100 (80GB VRAM) & 07:43  \\
        M10 & CodeLlama-13b &  A100 (80GB VRAM) & 14:21 \\
        \hline
    \end{tabular}
    \caption{Information about the runtime (minutes:seconds) of the experiments for generating cross-entropy of the 304 Java classes. We estimate the footprint of the experiments to be less than 0.15 kg CO2 eq. (estimation done with \url{https://mlco2.github.io/impact/}).}
    \label{tab:runtime}
\end{table*}

\section{Results}\label{sec:results}
Before diving into the results, it should be noted that the results for the model M7 (CodeGPT-java-10.2) are inverted with regard to the other models. Despite our investigations, we were not able to determine the reason of this inversion (the model has been run with the same code as the others). We report the results including M7 and highlights its impact on the results when necessary.

The other thing we notice is that larger models tend to have a lower and narrower range in terms of cross-entropy (see Table \ref{tab:desc-stats}).
\begin{table}[]
    \centering
    \begin{tabular}{lrrrrrrr}
    \hline
    ID & min. & max. & range & mean & var. & kurt. & skew. \\
    \hline
    M1 & 0.573 & 5.154 & 4.581 & 1.727 & 0.604 & 3.108 & 1.644 \\
    M2 & 0.408 & 4.659 & 4.252 & 1.456 & 0.554 & 2.469 & 1.609 \\
    M3 & 0.383 & 4.372 & 3.990 & 1.436 & 0.526 & 2.326 & 1.575 \\
    M4 & 0.376 & 4.518 & 4.143 & 1.385 & 0.582 & 2.558 & 1.695 \\
    M5 & 0.360 & 4.275 & 3.915 & 1.238 & 0.489 & 2.675 & 1.687 \\
    M6 & 1.156 & 8.466 & 7.310 & 4.267 & 1.406 & 1.126 & 0.800 \\
    M7 & 1.035 & 4.852 & 3.817 & 3.032 & 0.687 & -0.766 & -0.466 \\
    M8 & 0.403 & 4.674 & 4.271 & 1.433 & 0.536 & 3.029 & 1.693 \\
    M9 & 0.054 & 2.131 & 2.077 & 0.688 & 0.155 & 2.190 & 1.508 \\
    M10 & 0.052 & 2.063 & 2.011 & 0.602 & 0.102 & 2.667 & 1.389 \\
    \hline
    \end{tabular}
    \caption{Descriptive statistics for the cross-entropy generated by the 10 models on the benchmark data set. Notice that M7 is the only model having a negative Kurtosis}
    \label{tab:desc-stats}
\end{table}

\begin{figure*}[!htbp]
    \centering
    \begin{subfigure}[b]{1\textwidth}
        \includegraphics[width=1\textwidth]{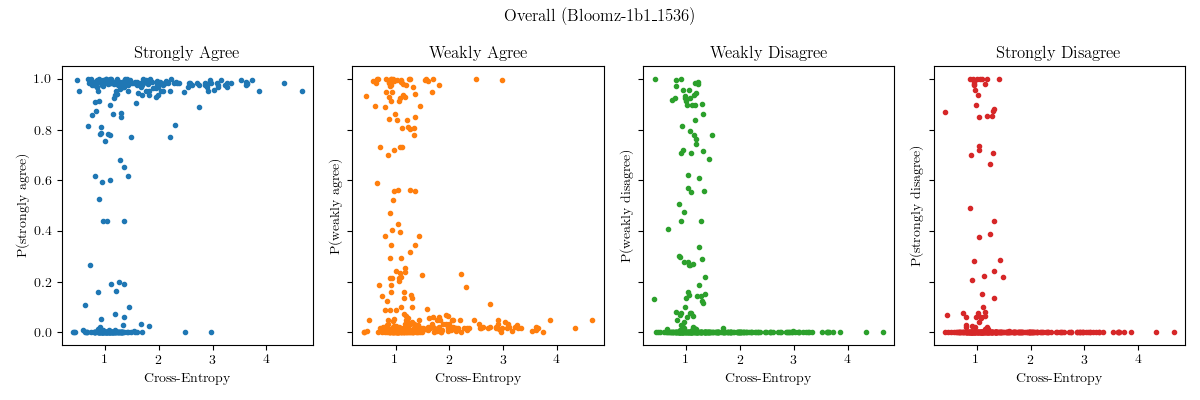}
        \caption{}
        \label{fig:example-association-a}
    \end{subfigure}
    \hfill
    \begin{subfigure}[b]{1\textwidth}
        \includegraphics[width=\textwidth]{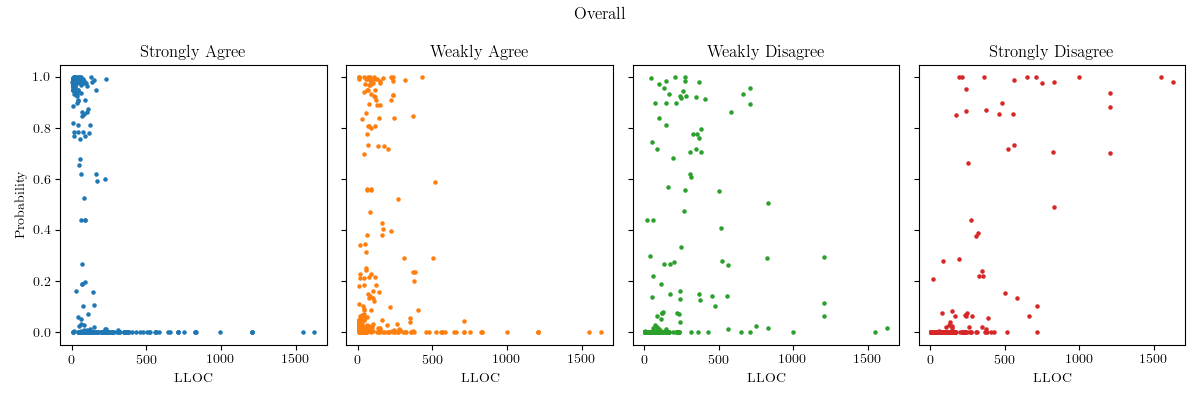}
        \caption{}
        \label{fig:example-association-b}
    \end{subfigure}
    \caption{Visualization of the relationship (a) between overall maintainability (Ov.) and cross-entropy (as computed by the model bloomz-1b1 (M2)); (b) between overall maintainability (Ov.) and LLOC.}
    \label{fig:example-association}
\end{figure*}

\begin{figure*}[!htbp]
    \centering
    \begin{subfigure}[b]{1\textwidth}
        \includegraphics[width=\textwidth]{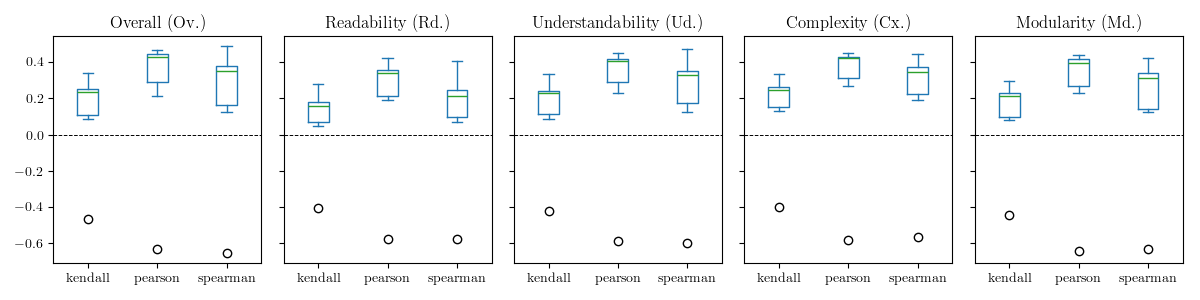}
        \caption{}
        \label{fig:assoc-all}
        \end{subfigure}
    \hfill
    \begin{subfigure}[b]{1\textwidth}
        \includegraphics[width=\textwidth]{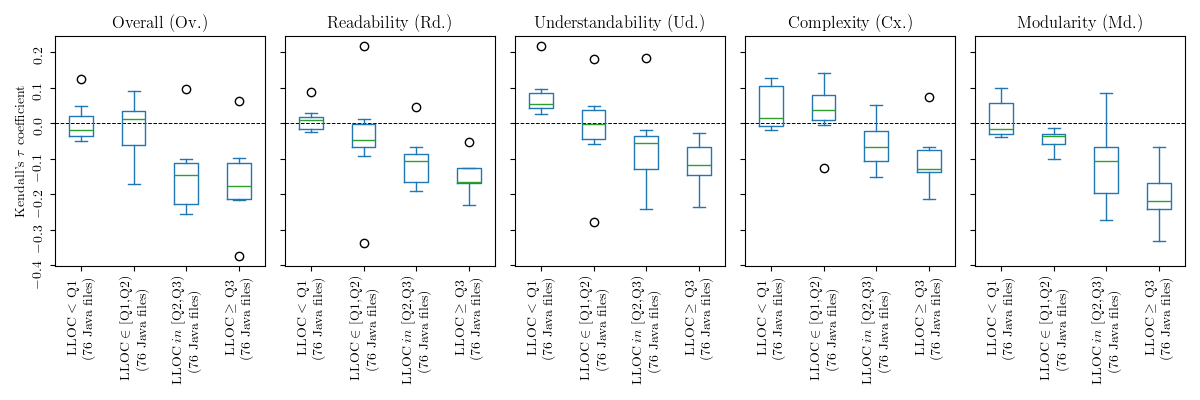}
        \caption{}
        \label{fig:assoc-lloc}
    \end{subfigure}
    \caption{Association between cross-entropy (measured by the 10 models) and the probability that experts would answer strongly agree for Ov., Rd., and Ud. and between cross-entropy and strongly disagree for Cx. and Md. (a) shows the association without stratification. (b) when stratifying by LLOC. Statistics for LLOC: min: 4, Q1: 17.0, Q2 (median): 56.5, Q3: 153.5, max: 1627.} 
\end{figure*}

Looking at the association between the cross-entropy generated by the 10 models and the probability of an expert answering "strongly agree" for each of the five dimensions (Figure \ref{fig:assoc-all}), we see a rather contradictory results. The association between cross-entropy and the probability of answering "strongly agree" are positive for overall maintainability (Ov.), readability (Rd.) and understandability (Ud.), and negative for complexity (Cx.) and modularity (Md.). This disagrees with the assumption that a higher cross-entropy is actually a sign of unexpected code and should be associated with a lower maintainability, i.e. the association should be negative for Ov., Rd., and Ud., and positive for Cx. and Md.

Rerunning the analysis while controlling for the number of logical lines of codes (LLOC) --- the number of lines of codes excluding comments and empty lines --- led to a reversal in the association between cross-entropy and the probability of "strongly agree" answer (see Figure \ref{fig:assoc-lloc} and Table \ref{tab:coef-logistic}) --- the association is near zero for small classes and stronger for larger classes. This is also highlighted by the coefficients of the logistic regression in table \ref{tab:coef-logistic}. This behaviour indicates a potential confounding effect from LLOC (see \cite{pearl2013simpson}) meaning that both the maintainability aspects and cross-entropy might be influenced by LLOC.

\begin{table*}[!htbp]
    \centering
    \small
    \begin{tabular}{cccccccc}
        \hline
        \multirow[c]{3}{*}{a)} & & coef & std err & z & P>|z| & [0.025 & 0.975] \\
        \cline{2-8}
         & const & 0.4120 & 0.134 & 3.082 & 0.002 & 0.150 & 0.674 \\
        & cross-entropy & 1.1517 & 0.172 & 6.677 & 0.000 & 0.814 & 1.490 \\
        \hline
        \hline
        \multirow[c]{3}{*}{b)} & & coef & std err & z & P>|z| & [0.025 & 0.975] \\
        \cline{2-8}
         & const & 0.5977 & 0.179 & 3.333 & 0.001 & 0.246 & 0.949 \\
         & LLOC & -2.7516 & 0.313 & -8.792 & 0.000 & -3.365 & -2.138 \\
        \hline
        \hline
        \multirow[c]{4}{*}{c)} & & coef & std err & z & P>|z| & [0.025 & 0.975] \\
        \cline{2-8}
        & const & 0.4018 & 0.191 & 2.104 & 0.035 & 0.028 & 0.776 \\
        & cross-entropy & -0.7963 & 0.306 & -2.602 & 0.009 & -1.396 & -0.196 \\
        & LLOC & -3.4271 & 0.444 & -7.725 & 0.000 & -4.297 & -2.558 \\
        \hline
    \end{tabular}
    \caption{Coefficients of the logistic regression while predicting P(stronlgy agree) > 0.5 for the overall maintainability (Ov.) for different inputs. Input features: a) log(LLOC), b) log(cross-entropy), c) log(LLOC) and log(cross-entropy). Cross-entropy is computed by model M2.}
    \label{tab:coef-logistic}
\end{table*}

\begin{figure*}[!htbp]
        \centering
        \includegraphics[width=1\textwidth]{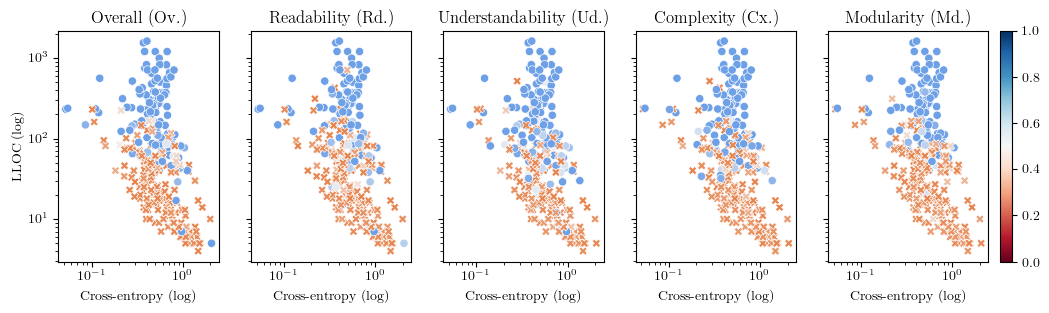}
    \caption{Visualisation of the feature space consisting of cross-entropy (x-axis, log scale) and LLOC (y-axis, log scale); 
    The colour scale represents the probability associated with the class (i.e. P(strongly agree) for Ov., Rd. and Ud. and P(strongly disagree) for Cx. and Md.).} 
    \label{fig:scatterplot}
\end{figure*}

Table \ref{tab:res-pred}, shows the results of predicting maintainability from either LLOC (baseline), cross-entropy, or both LLOC and cross-entropy. Figure \ref{fig:scatterplot} illustrates the input feature space for the classification task. Note that we have not found any significant differences in the prediction metrics between the 10 models (apart from M7 that have a reverse association in comparison to the other 9 models). Complexity of the LLM does not seem to play a significant role here. From the result in table \ref{tab:res-pred}, one can see that using solely cross-entropy is not as effective as using LLOC, and that using both LLOC and cross-entropy does not significantly improves the prediction metrics compared to the baseline.

\begin{table*}[!htbp]
    \centering
    \begin{tabular}{ccc}
        \hline
        \textbf{Dimension} & \textbf{\#0} & \textbf{\#1} \\
        \hline
        Overall (Ov.) & 130 & 174 \\
        Readability (Rd.) & 121 & 183 \\
        Understandability (Ud.) & 148 & 156 \\
        Complexity (Cx.) & 123 & 181 \\
        Modularity (Md.) & 109 & 195 \\
        \hline
    \end{tabular}
    \caption{Number of instances labeled 0 (\#0) or 1 (\#1) for each maintainability dimension}
    \label{tab:number-instances-labels}
\end{table*}

\begin{table*}[!htbp]
    \centering
    
    \begin{tabular}{ll|rrrr|rrrr|rrrr}
        \hline
         &  & \multicolumn{4}{c|}{Overall (Ov.)} & \multicolumn{4}{c|}{Readability (Rd.)} & \multicolumn{4}{c}{Understandability (Ud.)} \\
         &  & Acc & F1 & MCC & ROC & Acc & F1 & MCC & ROC & Acc & F1 & MCC & ROC \\
        \hline
        \multirow[t]{3}{*}{a)} & C1 & 0.842 & 0.860 & 0.683 & 0.838 & 0.796 & 0.835 & 0.578 & 0.780 & 0.799 & 0.801 & 0.605 & 0.799 \\
         & C2 & 0.758 & 0.792 & 0.514 & 0.748 & 0.748 & 0.797 & 0.473 & 0.729 & 0.695 & 0.717 & 0.394 & 0.693 \\
         & C3 & 0.835 & 0.853 & 0.670 & 0.833 & 0.789 & 0.838 & 0.554 & 0.762 & 0.792 & 0.798 & 0.591 & 0.792 \\
        \cline{1-14}
        \multirow[t]{3}{*}{b)} & C1 & 0.645 & 0.695 & 0.279 & 0.627 & 0.625 & 0.719 & 0.174 & 0.578 & 0.657 & 0.640 & 0.328 & 0.659 \\
         & C2 & 0.609 & 0.649 & 0.214 & 0.600 & 0.594 & 0.656 & 0.159 & 0.578 & 0.589 & 0.596 & 0.185 & 0.589 \\
         & C3 & 0.662 & 0.644 & 0.361 & 0.672 & 0.645 & 0.687 & 0.264 & 0.628 & 0.670 & 0.598 & 0.372 & 0.675 \\
        \cline{1-14}
        \multirow[t]{3}{*}{c)} & C1 & 0.847 & 0.865 & 0.698 & 0.843 & 0.791 & 0.828 & 0.574 & 0.775 & 0.805 & 0.811 & 0.619 & 0.804 \\
         & C2 & 0.795 & 0.819 & 0.592 & 0.790 & 0.749 & 0.792 & 0.481 & 0.733 & 0.726 & 0.734 & 0.458 & 0.726 \\
         & C3 & 0.834 & 0.855 & 0.672 & 0.830 & 0.788 & 0.830 & 0.561 & 0.765 & 0.797 & 0.806 & 0.601 & 0.796 \\
        \hline
    \end{tabular}
    \\ \vspace{2mm}

    \begin{tabular}{ll|rrrr|rrrr}
        \hline
         &  & \multicolumn{4}{c|}{Complexity (Cx.)} & \multicolumn{4}{c}{Moularity (Md.)} \\
         &  & Acc & F1 & MCC & ROC & Acc & F1 & MCC & ROC \\
        \hline
        \multirow[t]{3}{*}{a)} & C1 & 0.836 & 0.864 & 0.668 & 0.826 & 0.878 & 0.906 & 0.740 & 0.862 \\
         & C2 & 0.766 & 0.806 & 0.520 & 0.752 & 0.816 & 0.856 & 0.608 & 0.795 \\
         & C3 & 0.832 & 0.863 & 0.662 & 0.818 & 0.878 & 0.906 & 0.740 & 0.862 \\
        \cline{1-10}
        \multirow[t]{3}{*}{b)} & C1 & 0.642 & 0.720 & 0.233 & 0.610 & 0.646 & 0.751 & 0.159 & 0.570 \\
         & C2 & 0.618 & 0.669 & 0.219 & 0.603 & 0.654 & 0.725 & 0.260 & 0.624 \\
         & C3 & 0.677 & 0.700 & 0.366 & 0.677 & 0.705 & 0.766 & 0.348 & 0.673 \\
        \cline{1-10}
        \multirow[t]{3}{*}{c)} & C1 & 0.837 & 0.863 & 0.674 & 0.827 & 0.877 & 0.905 & 0.736 & 0.861 \\
         & C2 & 0.789 & 0.818 & 0.576 & 0.781 & 0.844 & 0.880 & 0.666 & 0.824 \\
         & C3 & 0.823 & 0.851 & 0.645 & 0.811 & 0.872 & 0.902 & 0.724 & 0.852 \\
        \hline
    \end{tabular}
    \caption{Results of the binary classification problem. Metrics: mean accuracy (Acc), area under the ROC curve (ROC), mean f1-score (F1), and mean Matthew Correlation Coefficient (MCC) computed over all 10 models (for each model the metrics are computed in a 10-fold cross validation setup). Classifiers: logistic regression without penalty (C1), random forest (C2), and support vector machine (C3). Input features: a) log(LLOC), b) log(cross-entropy), c) log(LLOC) and log(cross-entropy).}
    \label{tab:res-pred}
\end{table*}

\section{Comparison with other related work}
Schnappinger et al. \cite{schnappinger_preliminary_2022} evaluated BERT-based models, traditional machine learning models and even image-based models on the same data set. On the binary problem, they report that the SVM classifier (using TF-IDF of the code as input features) performs best, obtaining similar F1 and MCC values as our study.

Second, Hu et al. \cite{hu_maintainability_2023} created a sophisticated framework (DeepM) that integrates several types of neural networks using different input features (code metrics, text based and structure based features). They measured performance based on pair comparisons and recorded an average ranking accuracy (equivalent of the ROC AUC \cite{fawcett2006roc}) of 0.875. A value slightly greater than our method.

Sengamedu et al. \cite{sengamedu_neural_2022} evaluated the use of cross-entropy on different levels (token error, unnatural code, repetitive code, and long and complex code) by asking experts whether their proposed solution is useful or not. They report the following results (useful/not useful/not sure): token error (9/0/2), unnatural code (34/22/20), repetitive code (34/23/6), and long and complex code (18/3/6).

All three papers use different models and data sets. This makes direct comparisons difficult. However, some conclusions can be drawn. While the best F1 measure of our approach outperforms the deep learning models used by Schnappinger et al, the framework proposed by Hu et al appears to have an advantage. However, their multifaceted deep learning structure, underpinned by a plethora of extracted features, requires significant preprocessing overhead. In contrast, our methodology is characterised by its minimalist and streamlined approach. As Sengamedu et al. evaluated their work against expert opinion, we cannot directly compare our work with theirs without making assumptions about the usefulness of a maintainability classifier.

\section{Discussion}\label{sec:disccusions}

An initial intent of the study was to experiment with both text generation and fill mask models. However, as we proceeded further into implementation and initial testing, it became apparent that the fill mask models posed significant computational challenges. Due to the necessity to iterate over each token, mask it, and then engage the model, the runtime surged drastically. To provide context, using the Bloomz-560m model for text generation resulted in a runtime of 50 seconds, whereas the CodeBERTa-small-v1\footnote{\url{https://huggingface.co/huggingface/CodeBERTa-small-v1}} fill mask model consumed a staggering 57 minutes 46 seconds. This disproportionate increase in runtime, especially considering that Bloomz-560 has seven times the number of parameters compared to CodeBERTa-small-v1, led us to disqualify the fill mask models such as BERT-based models from our experiments, focusing our efforts primarily on text generation models based on GPT-2 and Llama-2.

At first, our strategy was to feed an entire Java class into the models. However, we encountered several problems. The tokenized representation of these classes often surpassed the maximum input size allowed by many models. Additionally, the memory footprint of the tokenization quickly outgrew the memory capabilities of the employed hardware, especially for tokenizers with a larger vocabulary.
Based on those findings, the idea emerged to divide a class into its constituent methods and subsequently compute cross-entropy as an average across these methods. This approach, however, presented two new challenges. First, accurately splitting the classes into methods while retaining their original formatting was difficult. Regular expressions were ill-suited to cover every possible scenario in a complex Java class, and using a parser to then rebuild the code from the Abstract Syntax Tree (AST) did not guarantee retention of the original format.
Second, despite the segmentation, individual methods could still breach the model's maximum input length.
Our final solution was to segment the entire class into chunks corresponding to the model's maximum permissible input size. This methodology, while pragmatic and effective, offers room for refinement in future endeavors.

\section{Threats to validity}\label{sec:treats_to_validity}
Validity of study results may be threatened by several factors associated with the study context and decision we made mainly due to time and resource constrains. Relevant threats to validity that should be considered while interpreting the study results and that should be addressed by future research include:

\textbf{Data validity}: the quality (correctness, consistency) of the maintainability assessments in the data set used in the study is described in \cite{schnappinger_dataset_2020}. Although the observed imbalance of the assessments was considered in the analysis, there might be other deficits – e.g., regarding correctness or consistency – that we cannot detect in the data but might have been potentially the case and thus influence the study outcomes.

\textbf{Selection bias}: due to time and resources constraints, an extensive large-scale comparison of multiple models and their parameters was not possible. Future research should extend the scope by other model types and settings. 

\textbf{Construct validity}: the average cross-entropy measure on a class level used in the study to compare generated and actual code might not be the most appropriate to quantify differences of code. Moreover, the assumption that LLM generate optimally maintainable code and thus can be used as a baseline might also be not true. This assumption should be verified, for instance, by letting human experts who assessed the maintainability of evaluation code, also assess the maintainability of the corresponding code generated by LLMs.

\textbf{Conclusion validity}: limited scope of the study – specific data and models – may have two consequences: (1) other types of models may outperform ours and thus led to other conclusions, e.g., that LLMs alone outperform models that involve measurement of code’s structural properties such as size (LLOC), (2) models created in the study may not apply to other software code, especially if other application domain or programming language is concerned.

\section {Conclusion}\label{sec:conclusion} 
In this paper we investigated the use of cross-entropy as a predictor of maintainability, approaching the problem from a "metric" point of view. Of course, LLMs have greater potential, as these tools can not only generate counterfactual code, but also explain it, and to some extent generate an argumentation as to why one piece of code is better than another. Our aim, of course, is not to reduce LLMs to code metrics calculators, but to shed light on the application of LLMs for the purpose of software quality management. 
Our study shows, like Sengamendu and Zhao \cite{sengamedu_neural_2022}, that, when controlling for the number of logical lines of code, cross-entropy is indeed associated with maintainability (the lower the cross-entropy, the higher the maintainability). This does not mean that LLMs can produce more maintainable classes, but that they have the potential to act as oracles for testing context-dependent and subjective aspects of code quality such as maintainability.
Our study left a number of open questions and thus room for future research. Examples of aspects to be investigated in the future include (1) the relationship with other code metrics, e.g. for software properties such as complexity, coupling or cohesion, (2) exploring the effect of using fine-tuned models, (3) verifying the assumption of baseline maintainability of the generated code, e.g. through assessment by subject matter experts. 

\section*{Acknowledgements}
This work is a result of the DeepQuali project, which has received funding from the KI4KU programme of the German Federal Ministry of Education and Research (BMBF) under grant agreement No 01IS23016D.

\bibliographystyle{plain}
\bibliography{main}

\begin{thebibliography}{10}

\bibitem{allamanis2019survey}
Miltiadis Allamanis, Earl~T. Barr, Premkumar Devanbu, and Charles Sutton.
\newblock A survey of machine learning for big code and naturalness.
\newblock {\em ACM Comput. Surv.}, 51(4), jul 2018.

\bibitem{briand2000exploring}
Lionel~C. Briand, Jürgen Wüst, John~W. Daly, and D.~{Victor Porter}.
\newblock Exploring the relationships between design measures and software
  quality in object-oriented systems.
\newblock {\em Journal of Systems and Software}, 51(3):245--273, 2000.

\bibitem{fawcett2006roc}
Tom Fawcett.
\newblock An introduction to roc analysis.
\newblock {\em Pattern Recognition Letters}, 27(8):861--874, 2006.
\newblock ROC Analysis in Pattern Recognition.

\bibitem{fenton1999critique}
Norman~E Fenton and Martin Neil.
\newblock A critique of software defect prediction models.
\newblock {\em IEEE Transactions on software engineering}, 25(5):675--689,
  1999.

\bibitem{hu_maintainability_2023}
Yamin Hu, Hao Jiang, and Zongyao Hu.
\newblock Measuring code maintainability with deep neural networks.
\newblock {\em Frontiers of Computer Science}, 17, 01 2023.

\bibitem{pearl2013simpson}
Judea Pearl.
\newblock Understanding simpson's paradox.
\newblock Technical report, Computer Science Department. University of
  California, Los Angeles, 2013.

\bibitem{schnappinger_defining_2020}
M.~Schnappinger, A.~Fietzke, and A.~Pretschner.
\newblock Defining a software maintainability dataset: Collecting, aggregating
  and analysing expert evaluations of software maintainability.
\newblock In {\em ICSME 2020. IEEE International Conference on Software
  Maintenance and Evolution}. IEEE, 2020.

\bibitem{schnappinger_dataset_2020}
Markus Schnappinger, Arnaud Fietzke, and Alexander Pretschner.
\newblock A software maintainability dataset, Sep 2020.

\bibitem{schnappinger_preliminary_2022}
Markus Schnappinger, Simon Zachau, Arnaud Fietzke, and Alexander Pretschner.
\newblock A preliminary study on using text- and image-based machine learning
  to predict software maintainability.
\newblock In Daniel Mendez, Manuel Wimmer, Dietmar Winkler, Stefan Biffl, and
  Johannes Bergsmann, editors, {\em Software Quality: The Next Big Thing in
  Software Engineering and Quality}, pages 41--60, Cham, 2022. Springer
  International Publishing.

\bibitem{sengamedu_neural_2022}
Srinivasan Sengamedu and Hangqi Zhao.
\newblock Neural language models for code quality identification.
\newblock In {\em Proceedings of the 6th {International} {Workshop} on
  {Machine} {Learning} {Techniques} for {Software} {Quality} {Evaluation}},
  pages 5--10. ACM, November 2022.

\bibitem{shepperd1988critique}
Martin Shepperd.
\newblock A critique of cyclomatic complexity as a software metric.
\newblock {\em Software Engineering Journal}, 3(2):30--36, 1988.

\bibitem{trendowicz2021data}
Adam Trendowicz, Julien Siebert, and Andreas Jedlitschka.
\newblock Data-driven technical debt management: Software engineering or data
  science challenge?
\newblock {\em IEEE Software}, 38(6):59--64, 2021.

\bibitem{wu2022survey}
Ruoting Wu, Yuxin Zhang, Qibiao Peng, Liang Chen, and Zibin Zheng.
\newblock A survey of deep learning models for structural code understanding.
\newblock {\em arXiv preprint arXiv:2205.01293}, 2022.

\bibitem{yang2022survey}
Yanming Yang, Xin Xia, David Lo, and John Grundy.
\newblock A survey on deep learning for software engineering.
\newblock {\em ACM Computing Surveys (CSUR)}, 54(10s):1--73, 2022.

\end{thebibliography}

\end{document}